\documentclass[traditabstract]{aa} 
\usepackage{graphicx}
\usepackage{txfonts}
\usepackage{natbib}
\usepackage{amssymb}
\begin{document}
%
   \title{Characteristics of the flare acceleration region derived from simultaneous hard X-ray and radio observations}


   \author{Hamish A. S. Reid\inst{1,2}, Nicole Vilmer\inst{2}, and Eduard P. Kontar\inst{1}}
   \institute{$^{1}$Department of Physics and Astronomy,
   University of Glasgow, G12 8QQ, United Kingdom\\
   $^{2}$ LESIA, Observatoire de Paris, CNRS, UPMC, Universit\'{e} Paris-Diderot, 5 place Jules Janssen, 92195 Meudon Cedex, France}

   \date{Received 22 Nov 2010 / Accepted 10 Feb 2011}

\abstract
{We investigate the type III radio bursts and X-ray signatures of accelerated electrons in a well observed solar flare in order to find the spatial properties of the acceleration region. Combining simultaneous RHESSI hard X-ray flare data and radio data from Phoenix-2 and the Nan\c{c}ay radioheliograph, the outward transport of flare accelerated electrons is analysed. The observations show that the starting frequencies of type III bursts are anti-correlated with the HXR spectral index of solar flare accelerated electrons. We demonstrate both analytically and numerically that the type III burst starting location is dependent upon the accelerated electron spectral index and the spatial acceleration region size, but weakly dependent on the density of energetic electrons for relatively intense electron beams. Using this relationship and the observed anti-correlation, we estimate the height and vertical extent of the acceleration region, giving values of around $50$~Mm and $10$~Mm, respectively. The inferred acceleration height and size suggest that electrons are accelerated well above the soft X-ray loop-top, which could be consistent with the electron acceleration between $40$~Mm and $60$~Mm above the flaring loop.}


\titlerunning{Simultaneous X-ray and radio observations}
\authorrunning{Reid et al}

   \maketitle


\section{Introduction}

Accelerated electron beams are believed to be responsible for both hard X-ray (HXR) and coherent radio emission during solar flares.  Upwards travelling electron beams propagate through the coronal plasma and sometimes escape into interplanetary space.  Emission from such beams is often observed as coronal and interplanetary type III radio bursts.  Electron beams travelling downwards with small pitch angles enter the dense plasma of the chromosphere and are generally seen through bremsstrahlung emission in HXR.  Before entering the chromosphere, downwards propagating electron beams may also produce reverse type III bursts.  Despite this wealth of electromagnetic beam emission from accelerated electrons propagating in plasma, the location of the electron acceleration site and its spatial characteristics are poorly known.

Indirect evidence of electron acceleration sites first came from broad band radio spectral observations of pairs of type III and reverse type III bursts \citep[e.g.][]{Aschwanden_etala1995,AschwandenBenz1997}.  The starting frequencies of these burst pairs are found between $220-910$~MHz, implying a range of electron densities in the acceleration region between $6\times10^8-10^{10}~\rm{cm}^{-3}$ for fundamental emission or $1.5\times 10^8 - 3\times 10^{9}~\rm{cm}^{-3}$ for harmonic emission. These densities are lower than ones observed in bright soft X-ray loops ($2\times 10^{10}- 2\times 10^{11}~\rm{cm}^{-3}$) suggesting that the acceleration region lies above the soft X-ray loops, being located for example in a cusp reconnection site.  HXR observations have also been independently used to provide insight into typical electron acceleration region heights above the photosphere.  Through electron time-of-flight analysis using HXR emission in the range 20-200~keV \citep{Aschwanden_etal1998}, height estimates have been found in the range 20-50~Mm.  The spatial size of the acceleration region still remains largely unknown.

The simultaneous observation of HXR and metric/decimetric radio emission is commonplace during flares and the relationship between type III bursts and hard X-ray emissions has been studied for many years \citep[see for example][for a review]{PickVilmer2008}. The first studies performed by \citet{Kane1972} found a good similarity between HXR and type III radio emission, suggesting the two emissions are produced by electrons originating from a common acceleration site.  Many subsequent studies have specifically dealt with the association of coherent type III radio emission and HXR bursts, both statistically \citep[e.g.][]{Kane1972,Kane1981,Hamilton_etal1990,Aschwanden_etala1995,ArznerBenz2005} and for individual events \citep[e.g.][]{Kane_etal1982,Benz_etal1983,Dennis_etal1984,Raoult_etal1985,Aschwanden_etalb1995,Raulin_etal2000,Vilmer_etal2002}.  A more recent statistical study of 201 flares above GOES class C5 \citep{Benz_etal2005} reports an $83\%$ association rate with coherent radio emission, within the range between 4~GHz and 100~MHz.  These results suggest that practically all flares with HXR GOES class $>$ C5 are associated with some form of coherent radio emission.

An in depth statistical study was carried out between radio type III bursts and HXRs by \citep{Kane1981}.  The study reported that the X-ray/type III correlation increases systematically with the intensity of HXR and radio emission, the peak spectral hardness of HXR emission and the type III burst starting frequency.  \citet{Hamilton_etal1990} similarly reported the systematic increase of HXR/type III correlation with increasing emission intensity and to a lesser extent with spectral index of HXR emission.  \citet{Hamilton_etal1990} also reported a statistical correlation between the peak HXR and type III intensities.   To produce a harder (smaller spectral index) HXR photon spectrum requires a harder electron beam spectrum.  A hard electron beam spectrum is an attractive attribute for type III producing electron beams as it makes it easier and faster for the bump-in-tail instability to occur.  Faster instability onset ties in very well with the HXR/type III correlation increasing for bursts with a higher starting frequency.

A temporal correlation between HXR and radio pulses has been found statistically \citep{Aschwanden_etala1995} where the average time delay betwen the HXR pulse and radio pulse starting frequency was $\le0.1$~s.  Temporal correlations have also been found in many individual event studies.  Of these studies the results by \citet{Dennis_etal1984} find a temporal correlation with a similar magnitude to the statistical study by \citet{Aschwanden_etala1995}.  This, together with previous correlations, suggests a common acceleration region with HXR producing electron beams having either slightly less distance to travel or slightly more energetic electrons responsible for the emission.  The simultaneous analysis of HXR and type III radio bursts is thus an attractive diagnostic of flare associated electron acceleration.

A few previous studies have attempted to infer properties (both height and size) of the common electron acceleration region from simultaneous HXR and radio observations.  \citet{Kane_etal1982} used an inferred spectral index from HXR emission to estimate the minimum distance required for the type III producing electron beam to become unstable.  With an assumed electron acceleration height of $20$~Mm, an altitude of $100$~Mm was deduced for the starting frequency, in good agreement with the spectral observations.  Unfortunately the type III frequencies in this analysis were too high with respect to the Nan\c{c}ay radioheliograph frequencies for radio imaging at this time so it was not possible to confirm the starting height of the radio emission.  \citet{Benz_etal1983} also considered the minimum distance required for a type III producing electron beam to become unstable.  By modelling both the HXR and radio producing electron beam as a Maxwellian they found a weak correlation between the type III starting frequency and the electron temperature derived from HXR observations above 26~keV.  However, in the event considered a change in electron temperature cannot fully account for the initial change in type III starting frequency.  The authors thus conclude that a movement of the acceleration site occurred for this event.

In another study \citet{Aschwanden_etala1995} uses the assumption of a common acceleration region producing upward and downward electron beams to estimate acceleration times and infer acceleration region sizes.  No frequency gap was observed between type III and reverse type III emission so their starting frequency separation distance was constrained by the detector resolution.  Using an assumed density model, this distance was found to be $<2$~Mm.  The instability distance for the electron beam was then equated to twice this resolution.  By using a similar analysis of \citet{Kane_etal1982} with observed HXR spectral index, acceleration times are found with $\Delta t < 0.3-3$~ms.  The size of the acceleration region is then inferred at $0.7$~Mm.  This constraint of acceleration site size and times is heavily dependent upon the assumption of bidirectional electron beams starting with a separation unresolved by the detector and thus should be treated with care.

Non-thermal HXR emission detected above 20~keV provides information on the overall spectrum of accelerated electrons, usually represented by a power-law in velocity and/or energy space.  It also provides insight on the density of the non-thermal electrons required to explain the HXR emission.  Radio type III bursts are believed to be formed via coherent plasma emission mechanisms requiring the presence of Langmuir wave turbulence \citep{GinzburgZhelezniakov1958}.  The high level of Langmuir waves infers a local electron beam-plasma instability.  Assuming that some acceleration mechanism in the corona produces a power-law in energy space the required unstable distribution can be formed due to electron transport parallel to the magnetic field.  Therefore, the criteria for beam instability and hence starting frequency of the type III emission are linked to the properties of the acceleration region; the spatial size, the spectral index of energised electrons, and the height in the corona.

In some events upward electron beams are able to travel from the corona to the inner heliosphere and create interplanetary type III bursts.  Previous work \citep{KontarReid2009,ReidKontar2010} has modelled electron beam transport from the Sun to the Earth taking into account the self-consistent generation of Langmuir waves.  \citet{KontarReid2009} found that wave-particle interactions are required to explain both electron beam onset times and spectral properties at 1~AU.  Developing this model further, \citet{ReidKontar2010} found the inclusion of density turbulence suppressed the growth of Langmuir waves which led to a broken power-law spectra similar to observational results \citep{Krucker_etal2009}.

Electron beams escaping into the inner heliosphere can also be detected in-situ near 1~AU and their numbers have been correlated to the number of HXR producing electrons \citep{Krucker_etal2007}. A correlation is found between the spectral indices of both electron populations as well as between the numbers of HXR producing electrons and escaping electrons for prompt electron events.  This again suggests that the X-ray producing electrons and the electrons moving upward in the corona originate from a common acceleration site. Furthermore,  it is found that the number of escaping electrons is on average only $∼0.2\%$ of the HXR-producing electrons above 50 keV.

In this paper we show that simultaneous radio and X-ray observations can be used with a simple analytical model to diagnose not only the location but the size of the acceleration region as well as the location.  This provides one of the first observational estimates of acceleration region size.  The present analysis of the April 15th 2002 at 08:51 UT suggests the acceleration region is located at heights of $\approx50$~Mm and has a spatial size of $\approx10$~Mm.  Numerical simulations are used with these inputs to help validate the results and explore unknown electron beam parameters.

\section{Basic concepts: starting frequency of type III bursts}

The aim of the following theory is to relate known observational variables from flares to unknown flare parameters.  Specifically we will relate the starting frequency of type III bursts and the spectral index of the inducing electron beam to the height and size of a flare acceleration region.  Initially let us consider the propagation of a flare accelerated electron cloud with starting characteristic size $d$ located at an initial height $r=0$.  The initial distribution is given by
\begin{eqnarray}\label{eq_init_f}
f(v,r,t=0)=g_0(v)exp(-|r|/d)
\end{eqnarray}
where $g_0(v)$ is the initial electron velocity distribution.  Langmuir waves will be generated when the quasilinear growth rate $\gamma(v,r)$ of electrons resonantly interacting ($\omega_{pe}(r)=kv$) with Langmuir waves \citep{Drummond_Pines1962, Vedenov_etal1962} becomes stronger than the collisional absorption $\gamma_c$ by the Maxwellian plasma

\begin{eqnarray}
\gamma(v,r) = \frac{\pi \omega_{pe}(r)}{n_e(r)}v^2\frac{\partial f}{\partial v}>\gamma_c,  \quad\quad\quad \gamma_c=\frac{\pi e^4 n_e(r)}{m_e^2v_{Te}^3}\ln \Lambda
\end{eqnarray}
where $\ln \Lambda $ is the Coulomb logarithm, taken near 20 for the parameters in the corona.  $\omega_{pe}(r)$, $n_e(r)$ and $v_{Te}$ are the background plasma frequency, density and thermal velocity respectively.

The initial velocity distribution of solar flare accelerated particles is usually taken as a power-law $g_0(v)\sim v^{-\alpha}$. This distribution at $t=0$ is stable and will not lead to generation of Langmuir waves. At later times $t>0$ the propagation of particles leads to the formation of a positive slope of the distribution function in velocity space. The distribution function changes in time due to the propagation (in the case of no energy losses)
\begin{eqnarray}
f(v,r,t)=g_0(v)exp(-|r-vt|/d)
\end{eqnarray}
and the growth rate for Langmuir waves becomes
\begin{eqnarray}
\gamma (v)=\frac{\pi \omega_{pe}(r)}{n_e(r)}v^2
f(v,r,t)\left(\frac{t}{d}-\frac{\alpha}{v}\right).
\label{eq_gamma}
\end{eqnarray}
We can observe strong Langmuir wave growth remotely via observations of type III radio emission.  Langmuir wave growth should occur at the distance $\Delta r=h_{typeIII}-h_{acc}$ from the original location where $h_{typeIII}$ is a height corresponding to the starting frequency of type III bursts  and $h_{acc}$ is the acceleration region height.  The distance $\Delta r$ can be found by equating the growth rate for Langmuir waves given in Equation (\ref{eq_gamma}) with $\gamma_c$ giving
\begin{eqnarray}
\Delta r=d\left(\alpha + \frac{\gamma_c n_e(r)}{\pi \omega_{pe}(r)}(v g_0(v))^{-1}\right).
\label{eq_x}
\end{eqnarray}
The quantity $vg_0(v)=n_b$ where $n_b$ is the inducing electron beam density.  By assuming a coronal background electron density of $10^9$~cm$^{-3}$, a coronal electron temperature of 2~MK and a beam density of $10^4$~cm$^{-3}$ the second term $\gamma_c n_e(r)/\pi \omega_{pe}(r)n_b$ is around $10^{-3} \ll \alpha$.  Thus the relation between the acceleration site properties and the type III starting frequency is determined mostly by the spectral slope of the electron beam giving the simple relation
\begin{eqnarray}
h_{typeIII}=d\alpha + h_{acc}
\label{eq_mx_c}
\end{eqnarray}
The unknown parameters $h_{acc}$ and $d$ can be found from the known parameters $\alpha$ and $h_{typeIII}$.  A key advantage of this relation lies with its lack of dependence on the poorly known electron beam number density necessary to produce type III emission.  The method is similar to what was discussed in \citet{Kane_etal1982}.  The key difference is that we consider an instantaneous injection at $t=0$ with a spatially broad distribution function while \citet{Kane_etal1982} considers a temporal injection from a point source.

\section{Observations and data analysis}
\label{ref:obs}

\subsection{Instruments}

In this study, RHESSI \citep{Lin_etal2002} data is used to analyse the HXR emission from downward propagating electron beams.  By assuming a common acceleration site for upward and downward electron beams, the X-ray emission observed by RHESSI can serve to provide the spectral index for the above prediction.  RHESSI can also provide spatial information regarding the HXR.

When observing the radio emission in the corona, the spectral information was obtained by using Phoenix-2 radiospectrometer \citep{Messmer_etal1999} within the frequency range 100 - 700~MHz.  This provides information we use to find the starting frequency of the type III radio emission.  The spatial information for type III radio emission was found by using the Nan\c{c}ay radioheliograph (NRH) \citep{KerdraonDelouis1997} in the frequency range of 164 - 432~MHz.  The spatial radio information observed by the NRH can make sure the observed radio flux is emitted from the same active region in the solar atmosphere responsible for the HXR emission.

\subsection{Selection}

We aimed our study at cases observed simultaneously with the NRH, Phoenix-2 and RHESSI.  We started our selection from a list of events presented in previous observational analysis between coherent radio and HXR emissions \citep{ArznerBenz2005,GrigisBenz2004,Benz_etal2005}.   Of the 58 events considered, 10 were found to have coherent radio emission in the frequencies covered by the NRH observations.  We selected one event which had a simple spatial configuration at all NRH radio wavelengths, clearly defined starting frequencies and a strong HXR flux.  This event was chosen as an illustration of the method.

\subsection{Measurements}

\begin{figure}
   \centering
\includegraphics[width=0.99\columnwidth]{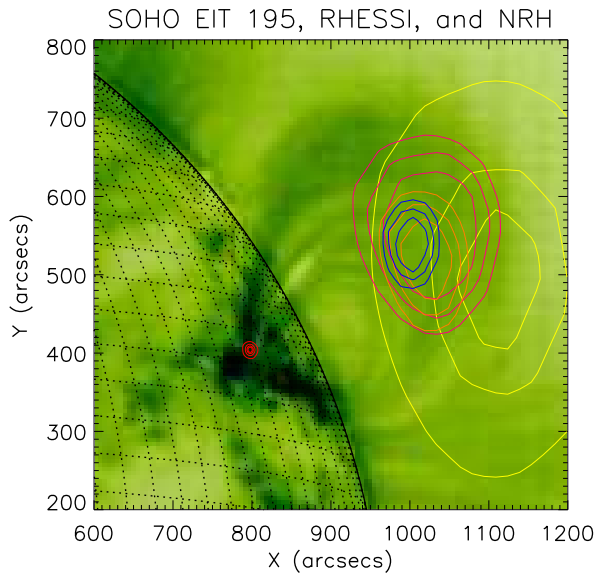}
   \caption{The morphology of the April 15, 2002 solar flare. Background is SOHO/EIT 195 image.  The small red contour lines at the base of the plasma loops on the left correspond to HXR photons imaged by RHESSI in the 15-30 keV range.  The large contours on the right hand side correspond to NRH radio images at frequencies 432~MHz (blue), 327~MHz (orange), 236~MHz (pink), 164~MHz (yellow).  }
 \label{fig:image}
\end{figure}

\begin{figure}
   \centering
\includegraphics[width=0.89\columnwidth]{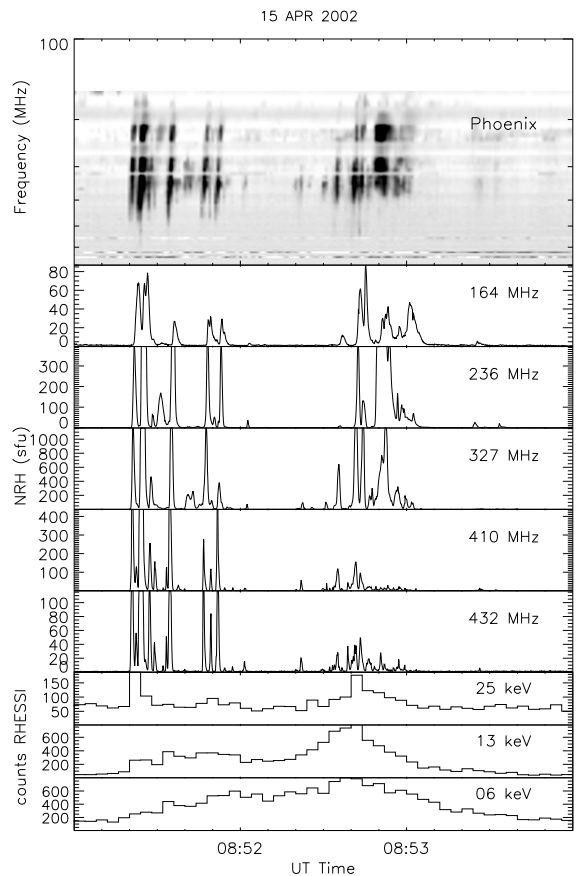}
   \caption{Time evolution of the radio and HXR fluxes for the April 15, 2002 solar flare between 08:51 and 08:54 UT.  The top panel is the Phoenix-2 radiospectrometer data on a log scale between the frequencies 160 and 700~MHz.  The middle panel is the Nan\c{c}ay radioheliograph flux time profiles observed at 5 discrete frequencies from 164 to 432~MHz.  The bottom panel are the RHESSI HXR counts/second at the three energy ranges 6-12, 12-25 and 25-50~keV.}
 \label{fig:spectr}
\end{figure}

The spatial overview of the April 15th flare is presented in Figure \ref{fig:image} using RHESSI, NRH and SOHO/EIT.  The X-ray source was imaged using RHESSI in the energy range between 15 and 30~keV. The higher energies had too few photons to make a reasonable image above the background noise.  The radio source was imaged using the NRH in frequency bands from 164 to 432~MHz and the size increases with decreasing frequency.  This can provide an estimate regarding the radial magnetic field expansion locally in the corona.  However, the decrease of spatial resolution with decreasing frequency using the NRH has to be considered.  This decrease behaves as $D/\lambda$ where D is the maximum antenna spatial separation and $\lambda$ is the wavelength of the radio emission.  The SOHO/EIT 195 image displays information about the overlying plasma configuration, conferring insight into the magnetic field structure where the flare originates.  The temporal evolution of the flare is presented in Figure \ref{fig:spectr}, using Phoenix-2, NRH and RHESSI data.

The spectral index of the X-ray emission, $\gamma$, was obtained using RHESSI spectral analysis of the photon flux $I(\epsilon) \sim \epsilon^{-\gamma}$.  The photon spectral index $\gamma$ was estimated using a power-law fit every 2 seconds (half-rotation of the spacecraft).  The one sigma error associated with the power-law fit was used as the spectral index error estimates.

The starting frequency of the type III radio emission was determined from the Phoenix-2 data.  We used Phoenix-2 data with a 1~sec temporal resolution.  The mean value of the radio flux on the quiet 3 minute interval 08:56 UT to 08:59 UT was used to quantify the background level for each frequency channel.  A threshold of twice this background level was then used at every moment in time to determine the starting frequency of the radio emission.  We then averaged the starting frequency over the 2 second RHESSI interval.  The mean width of the radio channels between 100 and 700 MHz was 9.2~MHz so we took 10~MHz as the one sigma error on the starting frequency.

The combined determination of starting frequencies and spectral indices was done on two time periods between 08:51:20 $\rightarrow$ 08:51:36 UT and 08:52:38 $\rightarrow$ 08:53:00 UT.  Both periods have a HXR non-thermal spectral index below 7.5 at all points in time.  Moreover, throughout both periods there is significant radio emission above the threshold frequency.

\begin{figure}
\centering
\includegraphics[width=0.99\columnwidth]{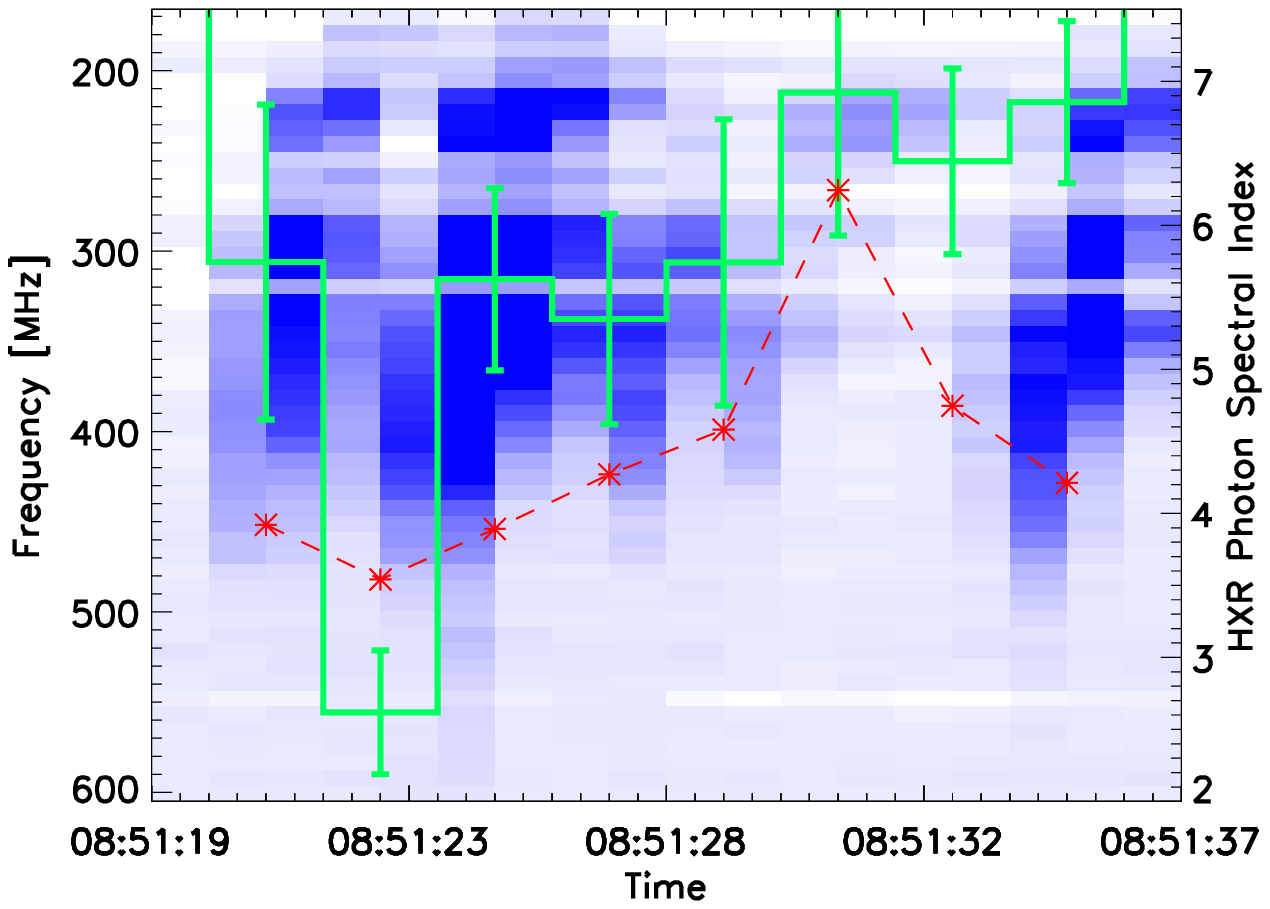}
\includegraphics[width=0.99\columnwidth]{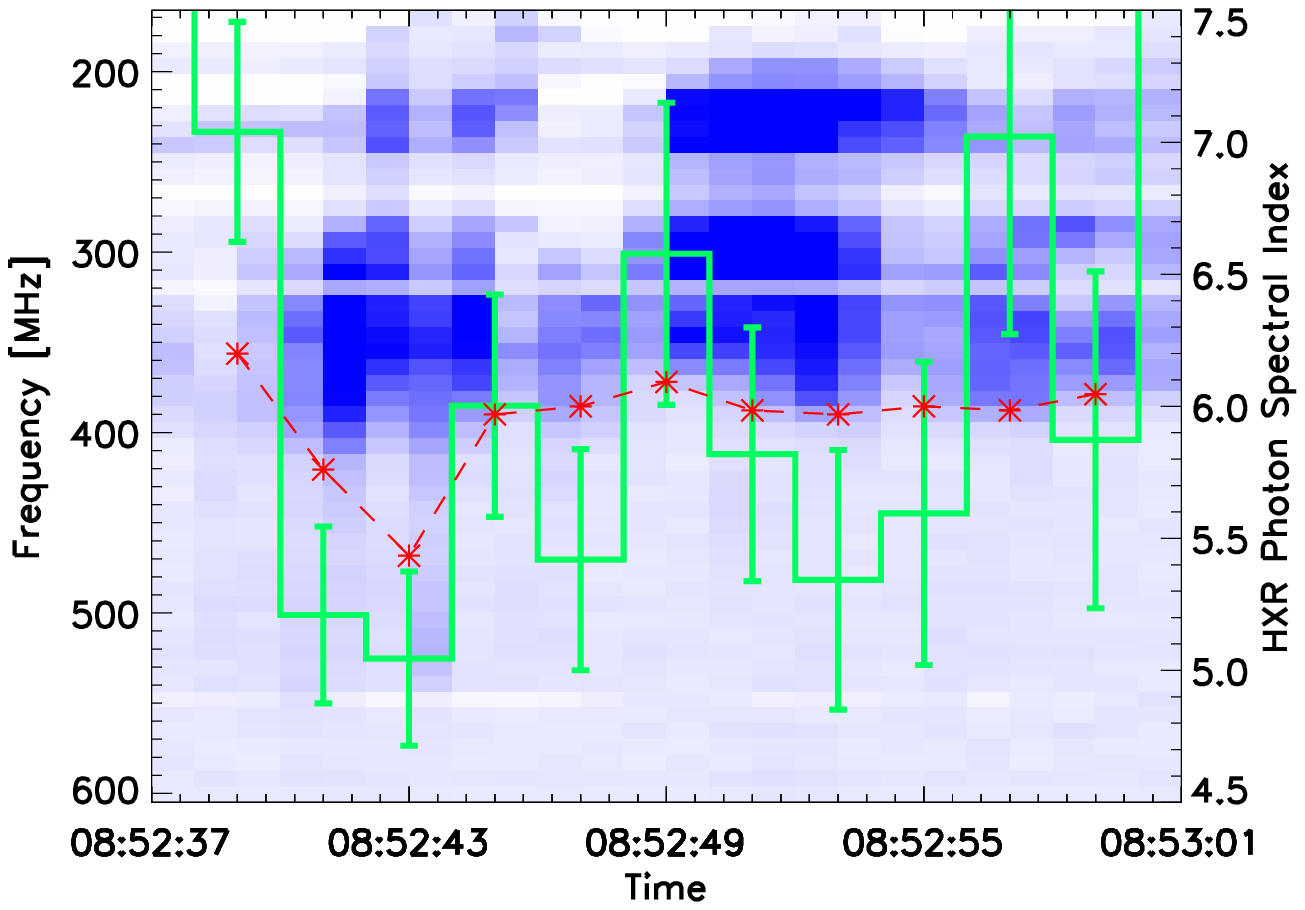}
\caption{HXR spectral index and frequency spectra of the type III burst for two different time periods in the April 15, 2002 event.  The starting frequencies are plotted as red stars connected by dashed lines.  The HXR spectral indices are plotted as 2 second green bars with error bars in the middle of their integration time.}
\label{fig:sp_index}
\end{figure}

\begin{figure}
\centering
\includegraphics[width=0.99\columnwidth]{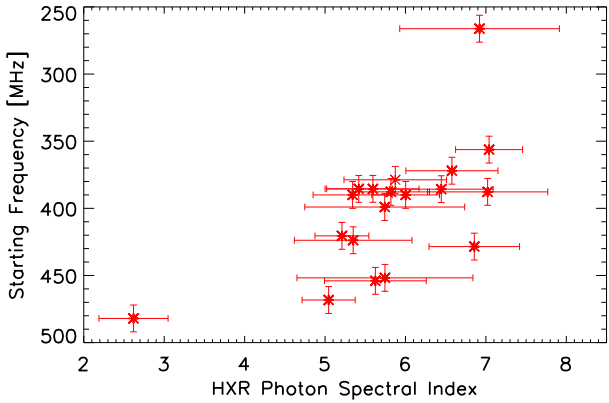}
\caption{Scatter plot of the HXR photon spectral index vs. the starting frequencies of the type III burst.  The one sigma observational errors on both spectral index and starting frequency are shown.}
\label{fig:SI_SF}
\end{figure}

The temporal evolution of $\gamma$ and the type III starting frequency is overplotted on the Phoenix-2 data in Figure \ref{fig:sp_index} for the time periods defined above.  The photon spectral index displays an anti-correlation with the type III starting frequency.  A clearer visualisation of this anti-correlation can be seen when both observables are plotted against each other (Figure \ref{fig:SI_SF}).  They have a Pearson correlation coefficient of -0.65 due to the starting frequency decreasing as the spectral index increases.  This correlation suggests that the two variables are related by a linear fit.

\subsection{Acceleration region}

To infer the characteristics of the coronal acceleration region from Equation (\ref{eq_mx_c}) we must use some assumptions to obtain $h_{typeIII}$ and $\alpha$ from the type III starting frequency and $\gamma$.  To relate the starting height of the type III emission $h_{typeIII}$ to the starting frequency we have used the exponential density model derived in \citet{Paesold_etal2001} which assumes second harmonic emission for a reference height of around $1.5$~$R_s$ for 160~MHz emission.   To obtain the electron beam spectral index in velocity space from the photon spectral index in energy space, the thick target model \citep{Brown1971, Brown_etal2006} was assumed.   The electron beam spectral index $\alpha$ can then be calculated from the photon spectral index $\gamma$ through the simple relationship $\alpha=2(\gamma+1)$.  The effect of photosphere albedo is ignored as the flare is located close to the limb \citep{KontarJeffrey2010}.

\begin{figure}
\centering
\includegraphics[width=0.99\columnwidth]{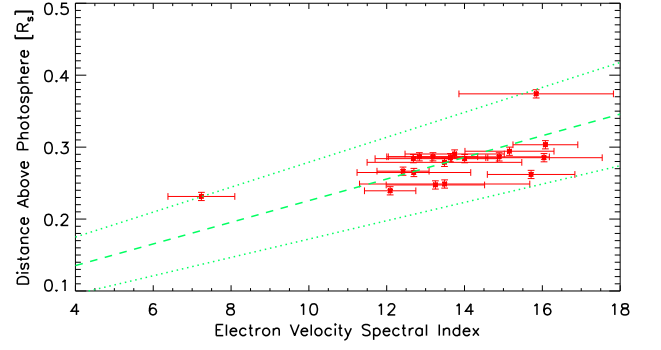}
\caption{Scatter plot of electron beam spectral index in velocity space, $\alpha$, calculated from the HXR photon spectral index vs. the distance above the photosphere associated with the starting frequencies of the type III burst.  The one sigma observational errors on both spectral index and height are shown.  The green dashed line is a linear fit to the data including observational error with the green dotted lines showing the extremes of the fit.}
\label{fig:sp_h}
\end{figure}

A positive correlation between the electron beam velocity spectral index and the starting height is observed with a Pearson correlation coefficient of 0.62 (Figure \ref{fig:sp_h}).  To investigate the correlation and obtain estimates of the acceleration region properties a linear fit was applied to the data.  The routine \emph{mpfitexy} \citep{Markwardt2009} was used to obtain a fit to the data including observational error (Figure \ref{fig:sp_h}).  Using Equation (\ref{eq_mx_c}) the linear fit infers the acceleration region height and size values of $h_{acc} = 52 \pm 21 $~Mm and $d=10.5 \pm 1.6 $~Mm respectively.  The larger percentage error of the height in relation to the size can be observed in the extremes of the fit shown in Figure \ref{fig:sp_h}.  If this linear relationship is statistically significant the slope has to be greater than zero.  We can test the null hypothesis that the slope is zero using the Students t-statistic \citep[e.g.][]{Num_Rec_C}.  A t-score of 6.56 is found with 19 degrees of freedom.  Using a confidence level of 0.01 we can comfortably reject the null hypothesis and say the linear relation is statistically significant.


The radio threshold frequency used to constrain the starting frequencies had a minor effect on the results if changed within reasonable parameters.  Different levels ($1.5 - 2.5 \times $ background level) changed the acceleration region properties by around $5-10~\%$.  Higher threshold frequencies caused higher $h_{acc}$ and lower $d$ with the converse being true.  Threshold frequencies $<1.5$ or $>2.5 \times $ background level caused unrealistic acceleration region parameters as starting frequencies were either not detected or were detected at high frequencies not corresponding to the visually observed type III emission.


\section{Numerical simulations of Langmuir wave generation}

To explore these predictions for acceleration height and size we can use numerical simulations of electron beams and induced Langmuir waves in the solar corona.  The simulations allow us to validate the observational result given the known initial conditions and check the validity of Equation (\ref{eq_mx_c}).  Moreover, it allows us to explore some of the unknown parameters such as beam density and the level of Langmuir waves required for radio emission.

\subsection{Electron beam transport}
The evolution of accelerated electrons can be considered self-consistently using weak turbulence theory where we have also taken into account binary collisions of energetic electrons with the surrounding plasma.  We consider the time evolution of an electron distribution function $f(v,r,t)$ and the induced Langmuir wave spectral energy density $W(v,r,t)$.  The energetic electrons travel in a beam-like structure along the magnetic field which radially expands outwards into the corona, allowing a one dimensional treatment of their propagation via
\begin{eqnarray}
\frac{\partial f}{\partial t} + \frac{v}{(r+r_0)^2}\frac{\partial}{\partial r}(r + r_0)^2f =
\frac{4\pi ^2e^2}{m^2}\frac{\partial }{\partial v}\frac{W}{v}\frac{\partial f}{\partial v} \quad\quad\quad\quad\quad\cr
	 +\frac{4\pi n_e(r) e^4}{m_e^2}\ln\Lambda\frac{\partial}{\partial v}\frac{f}{v^2}
\label{eqk1}
\end{eqnarray}
\begin{eqnarray}
\frac{\partial W}{\partial t} + \frac{\partial \omega_L}{\partial k}\frac{\partial W}{\partial r}
-\frac{\partial \omega _{pe}(r)}{\partial r}\frac{\partial W}{\partial k}
= \frac{\pi \omega_{pe}(r)}{n_e(r)}v^2W\frac{\partial f}{\partial v}
 \quad\quad\quad\quad\cr
- (\gamma_{c} +\gamma_L )W + e^2\omega_{pe}(r) v f \ln{\frac{v}{v_{Te}}}.
\label{eqk2}
\end{eqnarray}
The background plasma is assumed to be a Maxwellian distribution with thermal velocity $v_{Te}$, density $n_e(r)$ and plasma frequency $\omega_{pe}(r)$.  The background plasma damps through collisions both the electron beam (last term of Equation (\ref{eqk1})) and the Langmuir waves ($\gamma_c = \pi n_e(r) e^4\ln\Lambda/(m_e^2 v_{Te}^3)$) \citep[e.g.][]{Melrose11980}.  Coulomb collisions also spontaneously induce Langmuir waves (last term of Equation (\ref{eqk2})) \citep[e.g.][]{Hannah_etal2009}.  The first terms on the right hand side of Equations (\ref{eqk1},\ref{eqk2}) describe the resonant interaction $\omega_{pe}(r) = kv$ of electrons and Langmuir waves \citep{Drummond_Pines1962, Vedenov_etal1962}.  This resonant interaction is also responsible for Landau damping of Langmuir waves with phase velocities near $v_{Te}$, described through $\gamma_L = \sqrt{\pi/2}\omega_{pe}(r)(v/v_{Te})^3\exp({-v^2/2v_{Te}^2})$ \citep{LifshitzPitaevskii1981}.

As the density of the background plasma changes with distance from the Sun, Langmuir waves experience wave refraction.  This effect is described through the third term in Equation (\ref{eqk2}) where the level of background plasma inhomogeneity is characterised through $l=\omega_{pe}(r)^{-1}\partial\omega_{pe}(r)/\partial r$ \citep[e.g.][]{Ryutov1969,Kontar2001d}.

The spatial propagation of electron beam and Langmuir waves are described by the second term in Equations (\ref{eqk1},\ref{eqk2}).  Electron beam spatial propagation also includes the magnetic field expansion within the corona \citep{ReidKontar2010}.  This expansion in the corona is modelled through an expanding cone which has an angle $\theta$ and starts at length $r_0$ from the acceleration region.  The coronal expansion conserves the total number of electrons such that for scatter-free propagation, $\int(r+r_0)^2 n_e(r) dr = const$ along the magnetic field lines.

\subsection{Initial conditions}

The system of equations is solved numerically for an initially localised electron beam
\begin{eqnarray}
\label{initial_f}
f(v,r,t=0)=g_0(v)exp\left(-\frac{|r|}{d}\right),
\;\; g_0(v) = \frac{n_{b}(\alpha -1) }{v_{min}}\left(\frac{v_{min}}{v}\right)^{\alpha}.
\end{eqnarray}
The initial distribution is characterised by $d$ the acceleration region size, $v_{min}$ the minimum electron velocity considered, $\alpha$ the beam spectral index, and $n_b$ the density of the electron beam above $v_{min}$.  The background plasma absorbs Langmuir waves with phase velocity close to the thermal velocity via Landau damping therefore we set $v_{min}$ to $8v_{Te}$ and safely ignore velocities lower than $v_{min}$.

The thermal spectral energy density of Langmuir waves \citep[e.g.][]{Hannah_etal2009} is the initial condition given by
\begin{equation}\label{init_w}
W_{Th}(v,r,t=0) = \frac{k_BT_e}{4\pi^2}\frac{\omega_{pe}(r)^2}{v^2}\log\left(\frac{v}{v_{Te}}\right)
\end{equation}
where $T_e$ is the background plasma temperature set at 2~MK and $k_B$ is Boltzmann constant.  This is the expression for the thermal level of a Maxwellian plasma when collisions are weak.

The background density $n_e(r)$ was defined using the same exponential density model \citep{Paesold_etal2001} that was assumed in the previous section.

\subsubsection{Observational constraints}

The values derived from the observations in the previous section constrain some of the key input parameters for the simulations.  The starting height, $h_{acc}=52$~Mm which corresponds to a background density of $n_e=3\times 10^9$~cm$^{-3}$ using the exponential density model given in \citet{Paesold_etal2001}.  This gives a plasma frequency of $500$~MHz relating to second harmonic emission of $1000$~MHz.  The characteristic beam size $d=10.5$~Mm.  The HXR spectral index $\gamma$ is found from the RHESSI observations (Figure \ref{fig:sp_h}) which allows us to constrain the electron velocity spectral index as $6\le \alpha \le 16$.

The NRH images of the type III radio emission allow us to observe how the radio source increases with decreasing frequency.  Such an observation can provide information regarding the magnetic field expansion.  The size of the radio emission at 237~MHz is approximately twice the size of the radio emission at 432~MHz taken at $30\%$ of emission level.  This was measured around the two peak times of emission at 08:51:21 UT and 08:52:42 UT.  However, the wavelength  $\lambda$ is approximately twice as large at 237~MHz compared to 432~MHz and so the angular resolution of the NRH is increased by two.  Moreover, scattering by density inhomogeneities will increases the apparent size of the coronal radio source more at higher wavelengths \citep{Bastian1994}.  We thus in the present case \emph{cannot} observationally resolve any significant radial expansion of the magnetic field.  Such a scenario is equivalent to type III producing electron beams propagating along thin coronal structures as observed in \citet[e.g.][]{Trottet_etal1982,Pick_etal2009}, or having very small radial expansion of the magnetic field in the low corona.  The expansion is much smaller than what would be expected for the inner heliosphere, where the magnetic field expands as a cone with an angle around $40^o$ \citep[e.g.][]{Steinberg_etal1985}.

The density of electron beams responsible for type III emission is believed to be small with \citet{Krucker_etal2007} finding them $0.2\%$ of the density of the downward propagating electron beams responsible for HXR emission above 50~keV.  With an initial background density of $3\times 10^9$~cm$^{-3}$ providing the upper limit to the downward propagating electron beam, the upward propagating electron beam was injected with a density of $n_b=10^4$~cm$^{-3}$ above $11$~keV.  However, observations show time dependent intensities of HXR photons which is related to the density of the inducing beam.  Such results could indicate the potential need to consider a changing beam density.  We note that Equation (\ref{eq_mx_c}) is independent of the density of the electron beam.  The starting frequencies found from the upwardly propagating electron beam should thus be insensitive to rather large changes in beam density.

\subsection{Numerical Results}

A high level of Langmuir waves is required to induce type III emission.  We can estimate the starting height, $h_{typeIII}$, from the simulations through the ratio of Langmuir wave spectral energy density to its initial thermal level $W(v,r,t)/W_{Th}(v,r,t=0)$ or $W/W_{Th}$.  The first point in phase space when $W/W_{Th}$ exceeds a certain level can give us insight into how electron beams with different spectral indices become unstable.

\begin{figure}
\centering
\includegraphics[width=0.99\columnwidth]{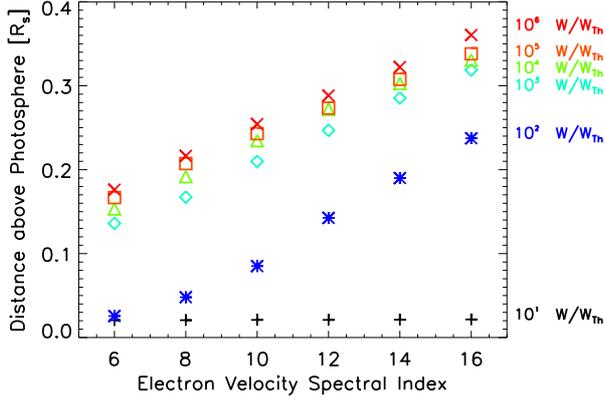}
\caption{Heights corresponding to high levels of Langmuir waves from an unstable beam with density $10^4~\rm{cm}^{-3}$.  Symbols and colours correspond to different levels of Langmuir wave growth.  Low levels (10~$W/W_{Th}$) correspond to spontaneous emission of waves.  High levels correspond to beam-plasma instability.}
\label{fig:num_delta_h}
\end{figure}

The numerical results are presented in Figure \ref{fig:num_delta_h} for a variety of different $W/W_{Th}$ levels.  By assuming Langmuir waves produce radio emission when they reach a certain level of $W/W_{Th}$, we can treat the curves in Figure \ref{fig:num_delta_h} in a similar manner to the observational results.  By applying a linear fit to each curve, we can obtain an estimate of the initial simulated acceleration region height and size using Equation (\ref{eq_mx_c}).  As we know the actual initial simulated values for $h_{acc}$ and $d$, these estimates allow us to check how accurate the method is for obtaining good estimates.  Such a fit also provides a numerical check for the simplified analytical relation Equation (\ref{eq_mx_c}) represents.  We find the closest fit to the simulated $h_{acc}$ and $d$ comes from the line where $W/W_{Th}=10^5$ giving $h_{acc}= 43.5\pm 5$~Mm and $d= 12.4\pm 0.6 $~Mm.  These variables are within $20\%$ and $15\%$ of the original numerical values respectively.


The results in Figure \ref{fig:num_delta_h} show a small variation between the heights corresponding to $10^3<W/W_{Th}<10^6$.  Provided there are enough electrons to generate sufficient Langmuir waves for radio emission, a change in the beam density has minimal effect on the starting height $h_{typeIII}$.  Increasing or decreasing the beam density by one order of magnitude changed the inducing height of $W/W_{Th}=10^5$ by at most $14~\%$ when $\alpha=6$ with a mean over all spectral indices of $3~\%$.  Changing the beam density will only vary the level of Langmuir waves which are produced upon the electron beam becoming unstable. This result confirms the density independence of Equation (\ref{eq_mx_c}) where instability of the electron beam is mainly dependent upon the spectral index and size of the electron cloud.

\begin{figure}
\centering
\includegraphics[width=0.99\columnwidth]{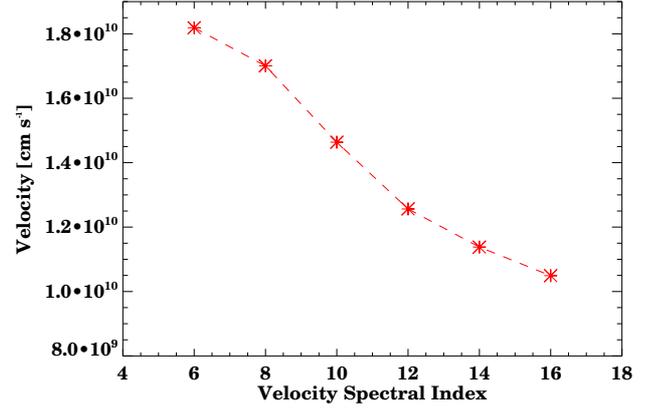}
\includegraphics[width=0.99\columnwidth]{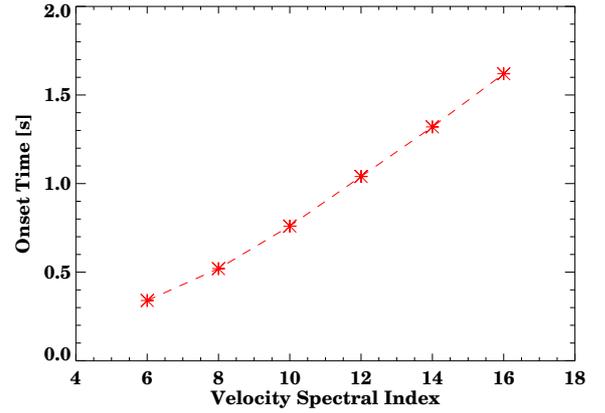}
\caption{Top: electron velocity spectral index plotted against the first phase velocity at which Langmuir waves exceed the threshold $W/W_{Th}=10^5$.  Bottom: electron velocity spectral index plotted against the onset time required for the electron beam to induce Langmuir waves exceeding the threshold $W/W_{Th}=10^5$.}
\label{fig:sp_vel_time}
\end{figure}

The ratio $W(v,r,t)/W_{Th}(v,r,t=0)$ also provides information regarding the Langmuir wave phase velocities and onset times when the waves exceeds certain thresholds.  As electrons resonantly interact with Langmuir waves, the phase velocity of the Langmuir waves conveys information regarding velocities of the inducing electrons.  The phase velocities corresponding to the points in Figure \ref{fig:num_delta_h} get smaller as the electron spectral index increases (softer spectrum).  An example of the velocity variation is presented in Figure \ref{fig:sp_vel_time} for the level $W/W_{Th}=10^5$.  Similarly the time required for the beam to induce Langmuir waves at a certain level increases for larger spectral index (Figure \ref{fig:sp_vel_time}).  Beams with larger spectral indices (with the same beam density) have less high energy electrons.  We thus expect Langmuir wave emission to be induced by lower energy electrons which take longer to become unstable.

We can explore how the inclusion of an expanding magnetic field would change the simulation results.  By ignoring the increase in radio source size at lower frequencies from NRH resolution and scattering we can assume the doubling of size between 432 and 237~MHz is because of a radially expanding magnetic field.  Using the assumed density model the magnetic field would then expand as a cone with an angle of $\theta = 6^o$.  By then assuming an acceleration site size which is as wide as it is long, we can constrain $r_0$ in Equation \ref{eqk1} to be $30$~Mm below the solar surface.  Running the simulations and using the same method described above gives estimations of $h_{acc}$ and $d$ which are $12\%$ and $40\%$ of their original values.  The expanding magnetic field causes the electron beam to induce a high level of Langmuir waves further away from the acceleration site.  The frequencies corresponding to these heights do not agree as well with the observed starting frequencies of the type III bursts.

\section{Discussion and conclusions}

We used simultaneous observations of radio and hard X-ray emission during a solar flare to gain insight about the acceleration region of energetic electrons.  With a simple model we have shown through an analytical relation how the starting height of type III emission and the spectral index of the electron beam can be related to the height and vertical extent of an acceleration region.  By combining HXR spectral information with the starting frequencies of the type III bursts, we have derived for our event an estimate on the acceleration site height and size of $h_{acc}=52 \pm 21 $~Mm and $d=10.5 \pm 1.6 $~Mm respectively.  We have also used self-consistent numerical simulations of an electron beam which can induce Langmuir waves in a background coronal plasma.  The simulations checked our predicted acceleration region values and allowed us to explore unknown parameters of the electron beam and Langmuir wave distributions.

The value found for $h_{acc}$ agrees with values in the range 20 - 50~Mm, deduced from electron time-of-flight analysis for HXR emission \citep{Aschwanden2002}.  This scenario indicates an acceleration region in the corona well above where SXR are imaged.  The error on $h_{acc}$ is quite large but within the 95~$\%$ range of 2-sigma the acceleration region remains within the corona.   The value found for $d$ is roughly an order of magnitude higher than previously found before in \citet{Aschwanden_etala1995}.  Assuming the relation in Equation (\ref{eq_mx_c}) the acceleration size in \citet{Aschwanden_etala1995} would not be able to produce significantly varying starting frequency of type III emission given a static acceleration site.  Such a small acceleration site predicted by \citet{Aschwanden_etala1995} may be relevant for type III radio bursts when very little or no evolution of the starting frequency can be observed.

Using the estimates for $h_{acc}$ and $d$ we ran self-consistent numerical simulations of an electron beam able to resonantly induce Langmuir waves in the background coronal plasma.  We analysed the distance required for a large magnitude of Langmuir waves to be induced through beam instability for a variety of initial beam spectral indices.  A linear fit to the initial beam spectral index and the height associated with strong Langmuir wave production gave a good estimation of the initial acceleration region height (within $20\%$) and the initial acceleration region size (within $15\%$) .  The result fits with the analytical predictions from Equation (\ref{eq_mx_c}) and hence the relation is a powerful diagnostic tool for flare acceleration site properties.  In line with the analytical equation the electron beam instability criteria was significantly dependent upon $\alpha$ and $d$ and almost independent on the beam density, which was confirmed by numerical simulations.  The simulations also gave an estimate of $W/W_{Th} \ge 10^5$ as the magnitude of Langmuir waves required to produce coherent radio emission.  The discrepancies found in the acceleration region properties are due to additional terms present in the numerical simulations which were not present for the simple analytical expression.

It is also possible to explore how a different assumed initial electron beam distribution in space will affect our results.  Initially in Equation \ref{eq_init_f} we assumed a tent distribution for the electrons in space.  We now consider an initial electron beam distribution which is Gaussian distributed in space such that
\begin{equation}
f(v,r,t=0)=g_0(v)exp(-r^2/d^2).
\label{eq_init_f2}
\end{equation}  The instability criteria for this distribution was already discussed in \citet{MelnikKontar1999} and, assuming small collisional damping, gives the relation
\begin{equation}
h_{typeIII} = 2d\sqrt{\alpha} + h_{acc}.
\label{eq_mx_c_2}
\end{equation}
The dependence of $h_{typeIII}$ on the square root of the spectral index originates from the $r^2$ in the exponential for the electron distribution function.  Unfortunately the observational errors on electron beam spectral index were too large to distinguish between the two models (Eq (\ref{eq_init_f}) and Eq (\ref{eq_init_f2})).  Even without observational error estimates such a fit to the data gives $h_{acc}=33 \pm 51$~Mm and $d=22\pm 7$~Mm which is not defined.  More detailed observations are thus needed to differentiate between different initial electron distributions in space.

Another assumption we considered was a static acceleration site during the entire event.  \citet{KaneRaoult1981} considered a moving acceleration site which decreased in altitude to explain why a type III burst's starting frequency increased with time.  Assuming the magnetic nature of reconnection, any source movement would typically be at the Alfven velocity.  At heights around $100$~Mm, this is typically around $1$~Mm s$^{-1}$ \citep{Arregui_etal2007}.  The Alfven velocity is too slow to account for the varying starting frequencies of the type III emission observed in the April 15th flare considered.  Moreover, the acceleration region would have to move upwards and downwards to account for the evolution of the starting frequencies.  Our results do not rule out the acceleration region moving in altitude but this will probably not be the dominant process for determining dynamic type III starting frequencies on a time scale of seconds.

Flares associated with the same active region responsible for the 08:51 UT flare on the April 15th 2002 have been analysed previously.  \citet{SuiHolman2003} found a coronal HXR source above the loop-top HXR source during another flare around 23:00 UT on the same day.  The high coronal HXR source was initially detected at an altitude of 25 Mm and moved with a speed of 0.3~Mm s$^{-1}$ up to an altitude of 40 Mm as the HXR flux increases. Moreover, the higher energy photons (15-16~keV) are detected at lower altitudes than the low energy photons (10-11~keV). This is indicative of an electron beam streaming down from a high acceleration region with high energy electrons having a larger stopping distance than low energy electrons.  Such a scenario fits with the derived high acceleration region $h_{acc} \approx 50$~Mm we found in this study.  A similar result was found for other high coronal sources \citep{Liu_etal2008,Liu_etal2009} where high energy photons are imaged at lower altitudes than low energy photons.

In conclusion, we stress that simultaneous HXR and radio observations are a tool to estimate the otherwise unmeasurable sizes of the acceleration site.  The results from our first trial of the relation given by Equation \ref{eq_mx_c} suggest that this size can be $\approx 10$~Mm located at height $\approx 50$~Mm, occupying a substantial fraction of the corona.  The size is larger than the HXR sources which are typically observed with RHESSI and in the range between a few Mm up to a few tens of Mm \citep{Emslie_etal2003}.  We hope to continue this avenue of research by extending it to more events to build a statistical picture of flare acceleration characteristics.  Future studies should also have a higher flux of HXR to better constrain the deduced electron beam spectral index.

\begin{acknowledgements}
Hamish Reid acknowledges the financial support of the SOLAIRE Network (MTRN-CT-2006-035484).  Nicole Vilmer acknowledges support from the Centre National d'Etudes Spatiales (CNES) and from the French program on Solar-Terrestrial Physics (PNST) of INSC/CNRS for the participation to the RHESSI project.  Eduard Kontar acknowledges financial support from a STFC rolling grant and STFC Advanced Fellowship. Financial support by the Royal Society grant (RG090411) is gratefully acknowledged. The overall effort has greatly benefited from support by a grant from the International Space Science Institute (ISSI) in Bern, Switzerland.  Collaborative work was supported by a British council Franco-British alliance grant.  The NRH is funded by the French Ministry of Education, the CNES and the R\'{e}gion Centre.
\end{acknowledgements}

\bibliographystyle{aa}
\bibliography{type3start_xray}

\begin{thebibliography}{50}
\expandafter\ifx\csname natexlab\endcsname\relax\def\natexlab#1{#1}\fi

\bibitem[{{Arregui} {et~al.}(2007){Arregui}, {Andries}, {Van Doorsselaere},
  {Goossens}, \& {Poedts}}]{Arregui_etal2007}
{Arregui}, I., {Andries}, J., {Van Doorsselaere}, T., {Goossens}, M., \&
  {Poedts}, S. 2007, \aap, 463, 333

\bibitem[{{Arzner} \& {Benz}(2005)}]{ArznerBenz2005}
{Arzner}, K. \& {Benz}, A.~O. 2005, \solphys, 231, 117

\bibitem[{{Aschwanden}(2002)}]{Aschwanden2002}
{Aschwanden}, M.~J. 2002, Space Science Reviews, 101, 1

\bibitem[{{Aschwanden} \& {Benz}(1997)}]{AschwandenBenz1997}
{Aschwanden}, M.~J. \& {Benz}, A.~O. 1997, \apj, 480, 825

\bibitem[{{Aschwanden} {et~al.}(1995{\natexlab{a}}){Aschwanden}, {Benz},
  {Dennis}, \& {Schwartz}}]{Aschwanden_etala1995}
{Aschwanden}, M.~J., {Benz}, A.~O., {Dennis}, B.~R., \& {Schwartz}, R.~A.
  1995{\natexlab{a}}, \apj, 455, 347

\bibitem[{{Aschwanden} {et~al.}(1995{\natexlab{b}}){Aschwanden}, {Montello},
  {Dennis}, \& {Benz}}]{Aschwanden_etalb1995}
{Aschwanden}, M.~J., {Montello}, M.~L., {Dennis}, B.~R., \& {Benz}, A.~O.
  1995{\natexlab{b}}, \apj, 440, 394

\bibitem[{{Aschwanden} {et~al.}(1998){Aschwanden}, {Schwartz}, \&
  {Dennis}}]{Aschwanden_etal1998}
{Aschwanden}, M.~J., {Schwartz}, R.~A., \& {Dennis}, B.~R. 1998, \apj, 502, 468

\bibitem[{{Bastian}(1994)}]{Bastian1994}
{Bastian}, T.~S. 1994, \apj, 426, 774

\bibitem[{{Benz} {et~al.}(1983){Benz}, {Barrow}, {Dennis}, {Pick}, {Raoult}, \&
  {Simnett}}]{Benz_etal1983}
{Benz}, A.~O., {Barrow}, C.~H., {Dennis}, B.~R., {et~al.} 1983, \solphys, 83,
  267

\bibitem[{{Benz} {et~al.}(2005){Benz}, {Grigis}, {Csillaghy}, \&
  {Saint-Hilaire}}]{Benz_etal2005}
{Benz}, A.~O., {Grigis}, P.~C., {Csillaghy}, A., \& {Saint-Hilaire}, P. 2005,
  \solphys, 226, 121

\bibitem[{{Brown}(1971)}]{Brown1971}
{Brown}, J.~C. 1971, \solphys, 18, 489

\bibitem[{{Brown} {et~al.}(2006){Brown}, {Emslie}, {Holman}, {Johns-Krull},
  {Kontar}, {Lin}, {Massone}, \& {Piana}}]{Brown_etal2006}
{Brown}, J.~C., {Emslie}, A.~G., {Holman}, G.~D., {et~al.} 2006, \apj, 643, 523

\bibitem[{{Dennis} {et~al.}(1984){Dennis}, {Benz}, {Ranieri}, \&
  {Simnett}}]{Dennis_etal1984}
{Dennis}, B.~R., {Benz}, A.~O., {Ranieri}, M., \& {Simnett}, G.~M. 1984,
  \solphys, 90, 383

\bibitem[{{Drummond} \& {Pines}(1962)}]{Drummond_Pines1962}
{Drummond}, W.~E. \& {Pines}, D. 1962, Nucl. Fusion Suppl., 3, 1049

\bibitem[{{Emslie} {et~al.}(2003){Emslie}, {Kontar}, {Krucker}, \&
  {Lin}}]{Emslie_etal2003}
{Emslie}, A.~G., {Kontar}, E.~P., {Krucker}, S., \& {Lin}, R.~P. 2003, \apjl,
  595, L107

\bibitem[{{Ginzburg} \& {Zhelezniakov}(1958)}]{GinzburgZhelezniakov1958}
{Ginzburg}, V.~L. \& {Zhelezniakov}, V.~V. 1958, Soviet Astronomy, 2, 653

\bibitem[{{Grigis} \& {Benz}(2004)}]{GrigisBenz2004}
{Grigis}, P.~C. \& {Benz}, A.~O. 2004, \aap, 426, 1093

\bibitem[{{Hamilton} {et~al.}(1990){Hamilton}, {Petrosian}, \&
  {Benz}}]{Hamilton_etal1990}
{Hamilton}, R.~J., {Petrosian}, V., \& {Benz}, A.~O. 1990, \apj, 358, 644

\bibitem[{{Hannah} {et~al.}(2009){Hannah}, {Kontar}, \&
  {Sirenko}}]{Hannah_etal2009}
{Hannah}, I.~G., {Kontar}, E.~P., \& {Sirenko}, O.~K. 2009, \apjl, 707, L45

\bibitem[{{Kane}(1972)}]{Kane1972}
{Kane}, S.~R. 1972, \solphys, 27, 174

\bibitem[{{Kane}(1981)}]{Kane1981}
{Kane}, S.~R. 1981, \apj, 247, 1113

\bibitem[{{Kane} {et~al.}(1982){Kane}, {Benz}, \& {Treumann}}]{Kane_etal1982}
{Kane}, S.~R., {Benz}, A.~O., \& {Treumann}, R.~A. 1982, \apj, 263, 423

\bibitem[{{Kane} \& {Raoult}(1981)}]{KaneRaoult1981}
{Kane}, S.~R. \& {Raoult}, A. 1981, \apjl, 248, L77

\bibitem[{{Kerdraon} \& {Delouis}(1997)}]{KerdraonDelouis1997}
{Kerdraon}, A. \& {Delouis}, J. 1997, in Coronal Physics from Radio and Space
  Observations, Vol. 483, 192--+

\bibitem[{{Kontar}(2001)}]{Kontar2001d}
{Kontar}, E.~P. 2001, \aap, 375, 629

\bibitem[{{Kontar} \& {Jeffrey}(2010)}]{KontarJeffrey2010}
{Kontar}, E.~P. \& {Jeffrey}, N.~L.~S. 2010, \aap, 513, L2+

\bibitem[{{Kontar} \& {Reid}(2009)}]{KontarReid2009}
{Kontar}, E.~P. \& {Reid}, H.~A.~S. 2009, \apjl, 695, L140

\bibitem[{{Krucker} {et~al.}(2007){Krucker}, {Kontar}, {Christe}, \&
  {Lin}}]{Krucker_etal2007}
{Krucker}, S., {Kontar}, E.~P., {Christe}, S., \& {Lin}, R.~P. 2007, \apjl,
  663, L109

\bibitem[{{Krucker} {et~al.}(2009){Krucker}, {Oakley}, \&
  {Lin}}]{Krucker_etal2009}
{Krucker}, S., {Oakley}, P.~H., \& {Lin}, R.~P. 2009, \apj, 691, 806

\bibitem[{{Lifshitz} \& {Pitaevskii}(1981)}]{LifshitzPitaevskii1981}
{Lifshitz}, E.~M. \& {Pitaevskii}, L.~P. 1981, {Physical kinetics}, ed.
  {Lifshitz, E.~M.~\& Pitaevskii, L.~P.}

\bibitem[{{Lin} {et~al.}(2002){Lin}, {Dennis}, {Hurford}, {Smith}, {Zehnder},
  {Harvey}, {Curtis}, {Pankow}, {Turin}, {Bester}, {Csillaghy}, {Lewis},
  {Madden}, {van Beek}, {Appleby}, {Raudorf}, {McTiernan}, {Ramaty}, {Schmahl},
  {Schwartz}, {Krucker}, {Abiad}, {Quinn}, {Berg}, {Hashii}, {Sterling},
  {Jackson}, {Pratt}, {Campbell}, {Malone}, {Landis}, {Barrington-Leigh},
  {Slassi-Sennou}, {Cork}, {Clark}, {Amato}, {Orwig}, {Boyle}, {Banks},
  {Shirey}, {Tolbert}, {Zarro}, {Snow}, {Thomsen}, {Henneck}, {McHedlishvili},
  {Ming}, {Fivian}, {Jordan}, {Wanner}, {Crubb}, {Preble}, {Matranga}, {Benz},
  {Hudson}, {Canfield}, {Holman}, {Crannell}, {Kosugi}, {Emslie}, {Vilmer},
  {Brown}, {Johns-Krull}, {Aschwanden}, {Metcalf}, \& {Conway}}]{Lin_etal2002}
{Lin}, R.~P., {Dennis}, B.~R., {Hurford}, G.~J., {et~al.} 2002, \solphys, 210,
  3

\bibitem[{{Liu} {et~al.}(2008){Liu}, {Petrosian}, {Dennis}, \&
  {Jiang}}]{Liu_etal2008}
{Liu}, W., {Petrosian}, V., {Dennis}, B.~R., \& {Jiang}, Y.~W. 2008, \apj, 676,
  704

\bibitem[{{Liu} {et~al.}(2009){Liu}, {Wang}, {Dennis}, \&
  {Holman}}]{Liu_etal2009}
{Liu}, W., {Wang}, T., {Dennis}, B.~R., \& {Holman}, G.~D. 2009, \apj, 698, 632

\bibitem[{{Markwardt}(2009)}]{Markwardt2009}
{Markwardt}, C.~B. 2009, in Astronomical Society of the Pacific Conference
  Series, Vol. 411, 251--+

\bibitem[{{Mel'Nik} \& {Kontar}(1999)}]{MelnikKontar1999}
{Mel'Nik}, V.~N. \& {Kontar}, E.~P. 1999, New Astronomy, 4, 41

\bibitem[{{Melrose}(1980)}]{Melrose11980}
{Melrose}, D.~B. 1980, {Plasma astrohysics. Nonthermal processes in diffuse
  magnetized plasmas} (Gordon and Breach)

\bibitem[{{Messmer} {et~al.}(1999){Messmer}, {Benz}, \&
  {Monstein}}]{Messmer_etal1999}
{Messmer}, P., {Benz}, A.~O., \& {Monstein}, C. 1999, \solphys, 187, 335

\bibitem[{{Paesold} {et~al.}(2001){Paesold}, {Benz}, {Klein}, \&
  {Vilmer}}]{Paesold_etal2001}
{Paesold}, G., {Benz}, A.~O., {Klein}, K., \& {Vilmer}, N. 2001, \aap, 371, 333

\bibitem[{{Pick} {et~al.}(2009){Pick}, {Kerdraon}, {Auch{\`e}re}, {Stenborg},
  {Bouteille}, \& {Soubri{\'e}}}]{Pick_etal2009}
{Pick}, M., {Kerdraon}, A., {Auch{\`e}re}, F., {et~al.} 2009, \solphys, 256,
  101

\bibitem[{{Pick} \& {Vilmer}(2008)}]{PickVilmer2008}
{Pick}, M. \& {Vilmer}, N. 2008, \aapr, 16, 1

\bibitem[{Press {et~al.}(1992)Press, Flannery, Teukolsky, \&
  Vetterling}]{Num_Rec_C}
Press, W., Flannery, B., Teukolsky, S., \& Vetterling, W. 1992, Numerical
  Recipes in C: The Art of Scientific Computing (Cambridge University Press)

\bibitem[{{Raoult} {et~al.}(1985){Raoult}, {Pick}, {Dennis}, \&
  {Kane}}]{Raoult_etal1985}
{Raoult}, A., {Pick}, M., {Dennis}, B.~R., \& {Kane}, S.~R. 1985, \apj, 299,
  1027

\bibitem[{{Raulin} {et~al.}(2000){Raulin}, {Vilmer}, {Trottet}, {Nitta},
  {Silva}, {Kaufmann}, {Correia}, \& {Magun}}]{Raulin_etal2000}
{Raulin}, J., {Vilmer}, N., {Trottet}, G., {et~al.} 2000, \aap, 355, 355

\bibitem[{{Reid} \& {Kontar}(2010)}]{ReidKontar2010}
{Reid}, H.~A.~S. \& {Kontar}, E.~P. 2010, \apj, 721, 864

\bibitem[{{Ryutov}(1969)}]{Ryutov1969}
{Ryutov}, D.~D. 1969, JETP, 57, 232

\bibitem[{{Steinberg} {et~al.}(1985){Steinberg}, {Hoang}, \&
  {Dulk}}]{Steinberg_etal1985}
{Steinberg}, J.~L., {Hoang}, S., \& {Dulk}, G.~A. 1985, \aap, 150, 205

\bibitem[{{Sui} \& {Holman}(2003)}]{SuiHolman2003}
{Sui}, L. \& {Holman}, G.~D. 2003, \apjl, 596, L251

\bibitem[{{Trottet} {et~al.}(1982){Trottet}, {Pick}, {House}, {Illing},
  {Sawyer}, \& {Wagner}}]{Trottet_etal1982}
{Trottet}, G., {Pick}, M., {House}, L., {et~al.} 1982, \aap, 111, 306

\bibitem[{{Vedenov} {et~al.}(1962){Vedenov}, {Lelikhov}, \&
  {Sagdeev}}]{Vedenov_etal1962}
{Vedenov}, A.~A., {Lelikhov}, E.~P., \& {Sagdeev}, R.~Z. 1962, Nucl. Fusion
  Suppl., 2, 465

\bibitem[{{Vilmer} {et~al.}(2002){Vilmer}, {Krucker}, {Lin}, \& {The Rhessi
  Team}}]{Vilmer_etal2002}
{Vilmer}, N., {Krucker}, S., {Lin}, R.~P., \& {The Rhessi Team}. 2002,
  \solphys, 210, 261

\end{thebibliography}

\end{document}